\documentclass[slac_one]{revtex4}
\usepackage{graphicx}
\usepackage{fancyhdr}
\usepackage{multirow}

\begin{document}

\title{Early Discoveries of New Gauge Bosons $W'$ and $Z'$ in Leptonic Decay Channels at ATLAS} 

%

\author{E.N. Thompson (for the ATLAS Collaboration)}
\affiliation{University of Massachusetts, Amherst, MA 01003}

\begin{abstract}
We present the potential of the ATLAS detector to discover new massive gauge bosons in their leptonic decay channels: $W'\rightarrow l^\pm\nu_l$ and $Z'\rightarrow l^+l^-$. Emphasis is placed on early data-taking at the LHC with low luminosity (up to 1~fb$^{-1}$).
\end{abstract}

\maketitle

\thispagestyle{fancy}


\section{MOTIVATION}
The Standard Model (SM) has thus far shown impressive predictive power and agreement with experiment.  However, there are some remaining puzzles which are not understood within the context of its present framework (for example, quadratic divergences to the radiative corrections for the Higgs mass, also referred to as the ``Hierarchy Problem'').  Beyond the Standard Model (BSM) theories, in the form of extended gauge symmetries, have been developed to address this and other incompleteness of the SM. Heavy charged and neutral gauge bosons, called the $W'$ and $Z'$, respectively, are a consequence of such theories, and may therefore appear around the TeV scale. The discovery potential of several of these BSM theories have been studied by the ATLAS collaboration.  In the most simple case, the Sequential Standard Model assumes quark and lepton couplings which are identical to the Standard Model, with cross sections scaled as $\left(M_{W,Z}/M_{W',Z'}\right)^2$ \cite{SSM}.  In another example, the Left-Right Symmetric Model arises as the SO(10) gauge group breaks to yield an extended SM gauge group consisting of SU(2)$_L$~$\times$~SU(2)$_R$~$\times$~U(1)~\cite{leftright}. As a consequence, this produces a right-handed charged boson ($W'_R$) as well as an additional neutral boson ($Z'$).  Leptonic decay channels ($W'\rightarrow l^\pm\nu_l$ and $Z'\rightarrow l^+l^-$) provide a clean signature for initial searches.  As of the date of this conference, the Tevatron and LEP experiments have set the current world limits on the $W'$ and $Z'$ masses just reaching 1 TeV for the various BSM theories \cite{PDG}.

\section{LEPTON SELECTION AND PERFORMANCE AT HIGH MOMENTUM}
The ATLAS detector is one of two general-purpose detectors on the LHC ring, and has been designed to reconstruct high-momentum particles in anticipation of new physics discoveries at the TeV scale.  We expect to have an integrated luminosity of up to 100~pb$^{-1}$ in the first year of data-taking, with a center of mass energy of 14~TeV \cite{detector}.  Background processes to $W'$ and $Z'$ production are greatly reduced in this energy regime, and the resonance (if it exists) would be a relatively clean signature.  Even so, calibration, alignment, and a general understanding of the performance of the detector are critical at the beginning of the experiment.  Misalignment in the Muon Spectrometer (MS), for example, can greatly reduce the significance of a 1~TeV $Z'$ resonance.  In preparation for data, tools are in place to determine performance quantities in situ, such as the electron and muon reconstruction efficiency, which can be determined via the ``tag and probe'' method with $Z\rightarrow e^+e^-$ and $\mu^+\mu^-$ events \cite{csc}.

%
%

\indent For the $W'$ and $Z'$ studies presented here, lepton candidates are chosen to be within the geometrical acceptance of the Inner Detector (pseudorapidity $|\eta| <$~2.5), and are required to have $p_T$ (transverse momentum) $>$ 50~GeV.  Electron candidates satisfy ``loose'' and ``medium'' selection criteria for $Z'$ and $W'$ searches, respectively.  The details of electron selection criteria can be found in \cite{csc}, but to summarize, the loose selection seeks to maximize efficiency by making only simple cuts on the shape of the calorimeter cluster, while the medium selection uses tighter requirements on the association between the position of the EM cluster and an ID track.  Muons at $\sim$500~GeV start to enter an energy regime where bremsstrahlung becomes more prevalent, which leads to poorer pattern recognition in the MS.  A $\chi^2 <$~100 requirement is made on track parameter matching between MS standalone tracks and associated ID tracks (for 5 degrees of freedom: two impact parameters, track $p_T$, $\eta$ and $\phi$).  For $W'\rightarrow l^\pm\nu_l$ studies, the neutrino is reconstructed by finding the total energy imbalance of the event in the transverse plane.  Details on missing transverse energy ($\displaystyle{\not}E_T$) reconstruction in a more general context can be found in \cite{csc}. The $\displaystyle{\not}E_T$ selection criteria is chosen to optimize resolution, and $W'$ candidates are required to have $\displaystyle{\not}E_T >$ 50~GeV.

\section{W' AND Z' EVENT SELECTION}

Though we are not able to fully reconstruct the $W'$ invariant mass without knowing the longitudinal momentum of the neutrino, using the transverse component of the missing energy it is possible to calculate the ``transverse mass'':
$$m_T = \sqrt{2p_T\displaystyle{\not}E_T\left(1-cos(\Delta\phi)\right)}$$
where $\Delta\phi$ is the angle between $p_T$ and $\displaystyle{\not}E_T^{jets}$ in the plane transverse to the beam axis. \\
\indent Dijet production is suppressed by using lepton isolation cuts and jet veto criteria, along with a lepton fraction requirement.  The lepton fraction in an event is defined as $\Sigma p_T^{leptons}/(\Sigma p_T^{leptons}+\Sigma E_T$), where $\Sigma p_T^{leptons}$ includes $\displaystyle{\not}E_T$ as well \cite{csc}.  The cross section for the various selection requirements can be found in Table \ref{Wcross}.

\begin{table}[ht]
\centering
\caption{Cross section for $m(W')$ = 1~TeV signal and backgrounds after each requirement in electron and muon channels \cite{csc}. \label{Wcross}}
\begin{tabular}{|l| c|c|c||c|c|c|}
\hline
\multirow{2}{*}{Requirement} & \multicolumn{3}{c||}{$\sigma$[pb] (electron channel)} & \multicolumn{3}{c|}{$\sigma$[pb] (muon channel)}\\
\cline{2-4}\cline{5-7}
 & $W' \rightarrow e^{\pm}\nu_e$ & $t\bar{t}$ & Dijets & $W'\rightarrow \mu^{\pm}\nu_{\mu}$ & $t\bar{t}$ & Dijets \\
\hline
(No requirement) & 4.99 & 452 & $1.91 \times 10^{10}$ & 4.99 & 452 & $1.91 \times10^{10}$\\

Preselection & 3.67$\pm$0.04 & 150.57$\pm$0.40 & (13.6$\pm$0.2) $\times 10^6$ & 4.28$\pm$0.05 & 205.30$\pm$0.46 & (11.2$\pm$0.19) $\times 10^6$\\

$p_T >$ 50 GeV & 3.43$\pm$0.04 & 51.13$\pm$0.23 & (7.23$\pm$0.6) $\times 10^3$ & 4.03$\pm$0.04 & 61.71$\pm$0.25 & (1.24$\pm$0.26)$\times 10^3$ \\

$E_T^{miss} >$ 50 GeV & 4.00$\pm$0.04 & 25.78$\pm$0.16 & 45$\pm$16 & 4.00$\pm$0.04 & 31.34$\pm$0.18 & 74$\pm$23\\

Lepton Isolation & 3.36$\pm$0.04 & 23.30$\pm$0.16 & 0.65$\pm$0.13 & 3.95$\pm$0.04 & 28.70$\pm$0.17 & 1.00$\pm$0.82 \\

Lepton fraction & 3.25$\pm$0.04 & 0.50$\pm$0.02& 0 & 3.81$\pm$0.04 & 0.64$\pm$0.03 & (1.96$\pm$1.38) $\times 10^{-3}$ \\
$m_T >$ 700 GeV & 1.86$\pm$0.03 & 0 & 0 & 2.20$\pm$0.03 & 0.007$\pm$0.003 & 0.001$\pm$0.001\\
\hline
\end{tabular}
\end{table}

\begin{figure}[th]
\begin{minipage}[b]{0.8\linewidth}
\includegraphics[width=2.2 in]{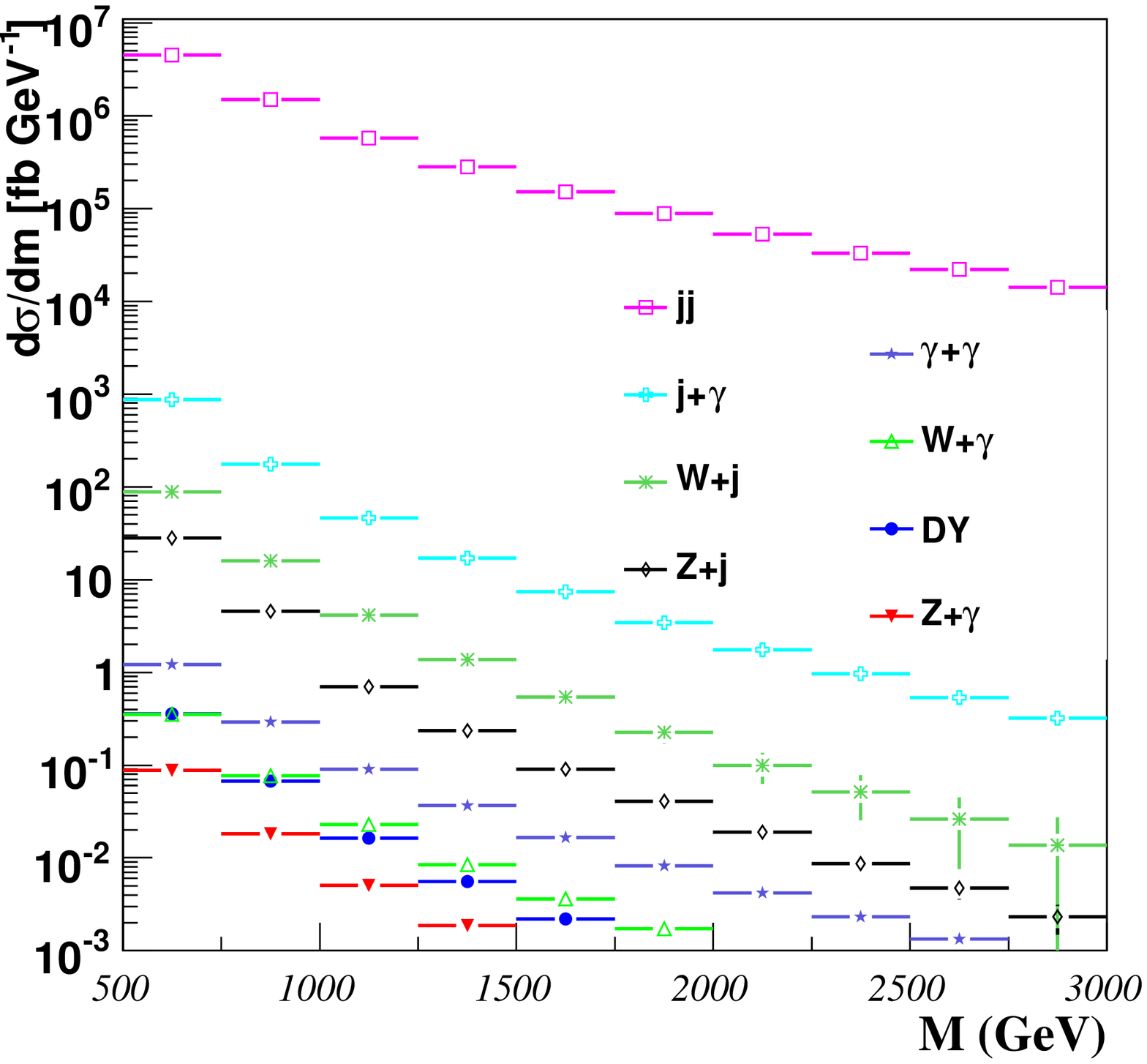}
\hspace{1.0cm}
\includegraphics[width=2.2 in]{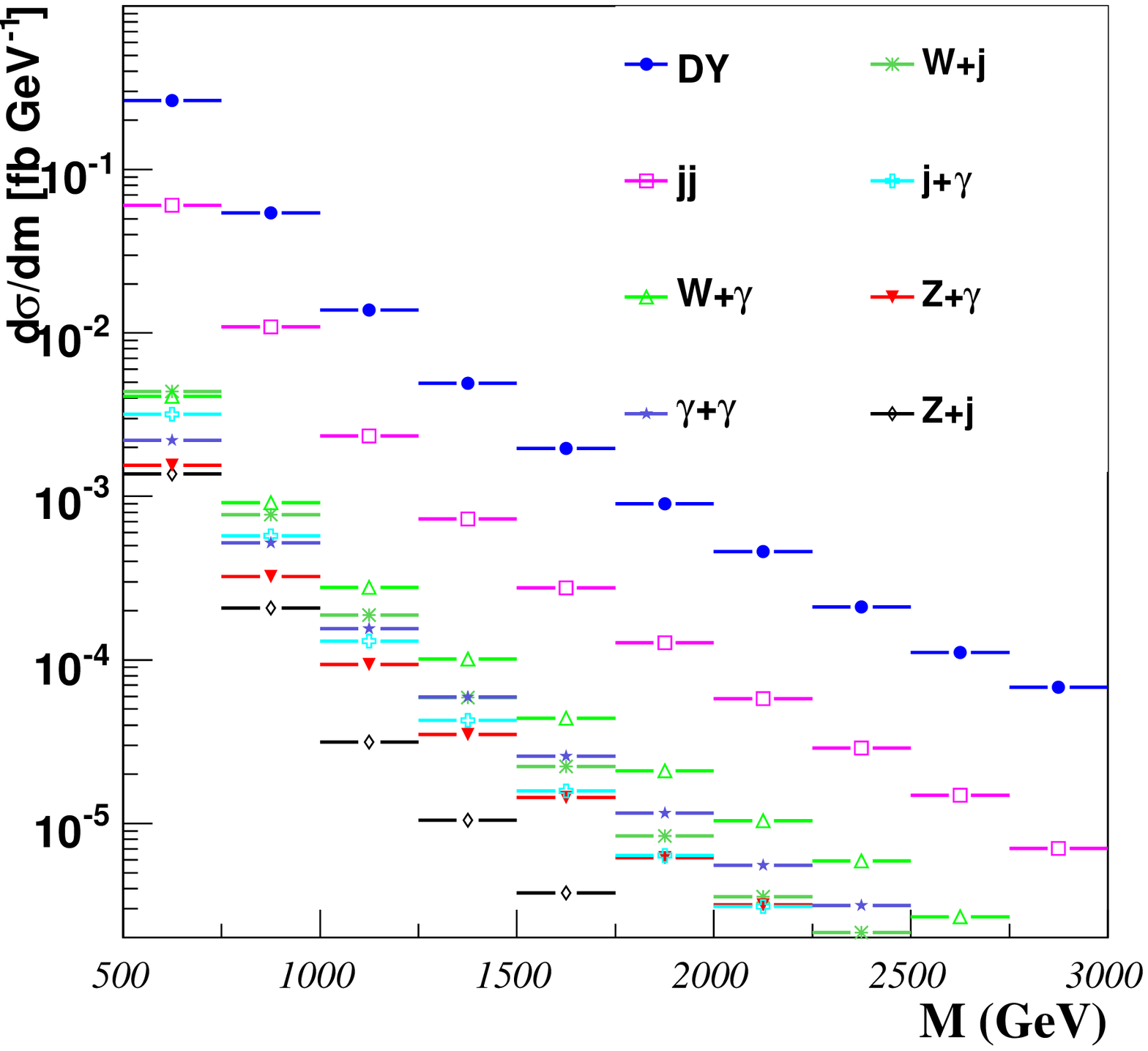}
\caption{Background contributions from various sources to the dielectron invariant mass spectrum before (left) and after (right) selection requirements \cite{csc}.}
\label{background}
\end{minipage}
\end{figure}

The $Z'\rightarrow l^+l^-$ resonance is also expected to have low background largely coming from $t\bar{t}$ and $b\bar{b}$ events. These contributions to the background can mainly be reduced by requiring isolated, oppositely-charged leptons (Figure \ref{background}).  Standard Model Drell-Yan production will be the principal irreducible background, but is expected to be only $\sim$1\% of a signal at 1~TeV \cite{csc}.

\section{SIGNAL EXTRACTION, SIGNIFICANCE AND EARLY DATA PROJECTIONS}

\begin{figure}[ht]
\begin{minipage}[b]{0.4\linewidth}
\includegraphics[width=2.8 in]{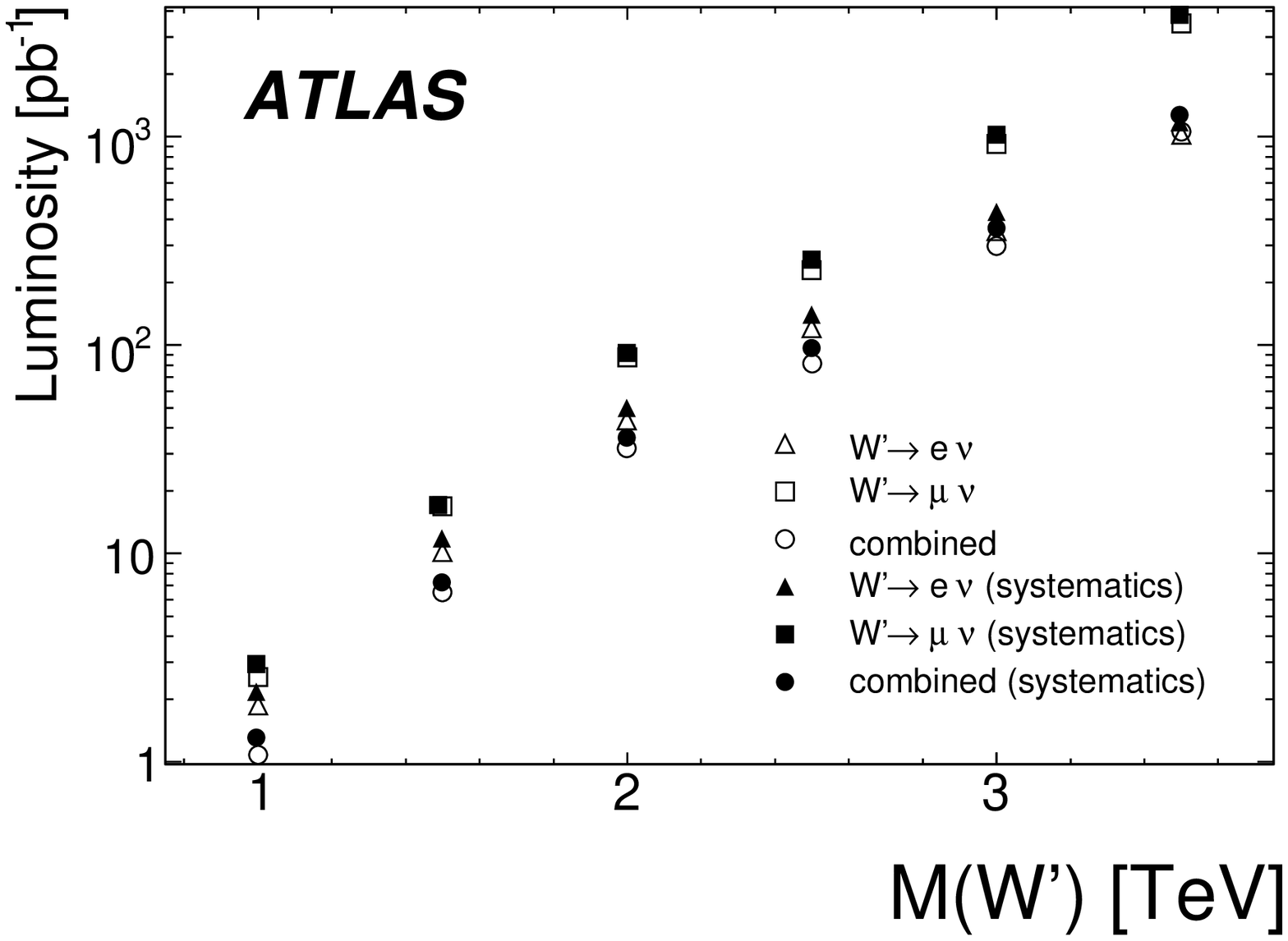}
\caption{Integrated luminosity needed for 5$\sigma$ discovery as a function of $m(W'_{SSM})$ \cite{csc}.}
\label{lepMETdisc}
\end{minipage}%
\hspace{0.6cm}
\begin{minipage}[b]{0.4\linewidth}
\includegraphics[width=2.35 in]{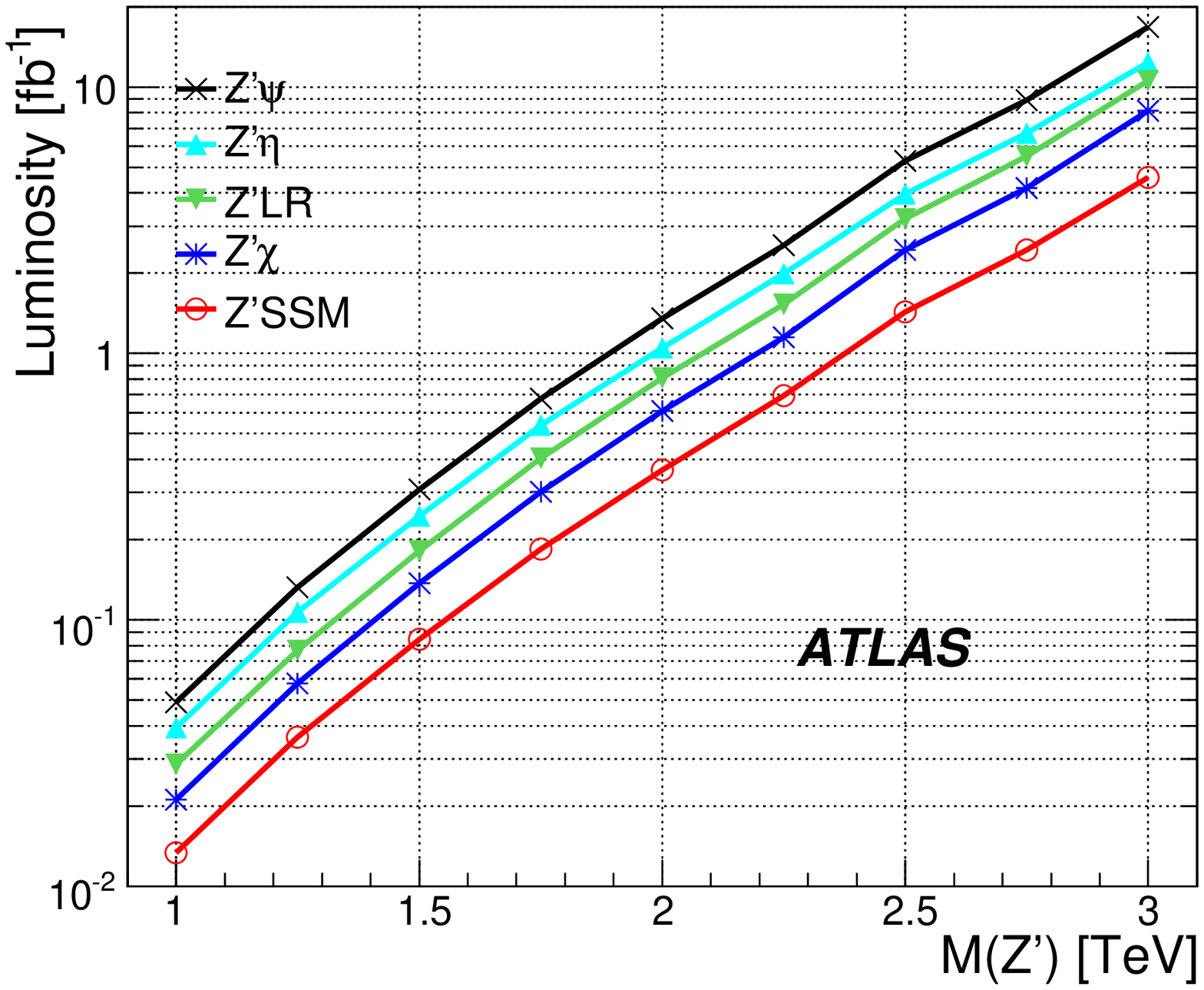}
\caption{
Integrated luminosity needed for a 5$\sigma$ discovery potential as a function of $m(Z')$ in dielectron channel, for various models
\cite{csc}.}
\label{dilepdisc}
\end{minipage}
\end{figure}

Extracting significance can be done in several ways. A number counting technique can be applied to set limits, where an estimator such as the log-likelihood ratio (LLH) is used to discriminate between a true signal and a statistical fluctuation of the background.  For this study, a fast Fourier transform (FFT) method was used to determine the LLH more quickly \cite{FFT}.  Sensitivity is much higher in the ``shape analysis'' approach, where each bin in a well known invariant mass spectrum is treated as a separate search. Data in the shape analysis approach is matched to a background only hypothesis (H0) and a signal plus background hypothesis (H1).  Confidence levels (CL) are then computed from the LLH ratio of the two models after many ensemble tests, and significance is given by the following (in the ``double tail convention''):
$$S = \sqrt{2}\times \mathrm{Erf}^{-1}\left(1-\frac{1}{CL_s}\right)$$
where $CL_s = CL_{H1}/CL_{H0}$ in the ``modified frequentist approach'' \cite{mod}.
The systematics included in these analyzes are estimated from Monte Carlo studies, but are expected to be conservative. Discovery potential curves are shown in Figures \ref{lepMETdisc} and \ref{dilepdisc}. Including systematics, a 5$\sigma$ $W'$ with mass of 1 TeV would be seen in tens of pb$^{-1}$, or $m(W') =$ 2 TeV in about 100 pb$^{-1}$ \cite{csc}. In the $Z'$ search, 20 to 40 pb$^{-1}$ would be needed to see a $m(Z') =$ 1 TeV resonance with 5$\sigma$, or about 1 fb$^{-1}$ for $m(Z') =$ 2 TeV \cite{csc}.

\indent Initial searches are mainly focused on discovery, but it is still pertinent to discuss discrimination between the various models.  For example, angular distributions of the decay products can discriminate between spin-1 and spin-2 resonances, the latter being expected in the case of graviton production. In addition, the lepton forward-backward asymmetry provides discrimination between various models, which have different boson-fermion couplings.  Eventually, as the detector is tuned and resolution is optimized, the product of the natural width of the resonance and its cross section can be used as another discriminant, although this is not expected to be possible in the first year of data taking.  Even so, early data will be used to gain an extensive understanding of the detector performance and SM background. Within the first year of running, ATLAS will extend the reach for $W'$ and $Z'$ particles beyond the current limits.


\end{document}